\documentclass[pre,twocolumn,superscriptaddress,floatfix]{revtex4-1}  

\usepackage[paperwidth=210mm,paperheight=297mm,centering,hmargin=2cm,vmargin=2.5cm]{geometry}

\usepackage[pdftex]{graphicx}
\usepackage{amsmath,amsfonts,amssymb}
\usepackage{natbib}
\usepackage{MnSymbol}
\usepackage{csquotes}

\usepackage[]{algorithm}
\usepackage{algorithmic}

\usepackage{hyperref}
\usepackage{xcolor} 

\definecolor{bluemoi}{rgb}{0.25,0.50 ,0.75} 

\hypersetup{
  bookmarksopen=true,
  pdftitle=" ",
  pdfauthor=" ", 
  pdfsubject=" ", 
  pdftoolbar=true, 
  pdfmenubar=true, 
  pdfhighlight=/O, 
  colorlinks=true, 
  pdfpagemode=UseNone, 
  pdfpagelayout=SinglePage, 
  pdffitwindow=true, 
  linkcolor=bluemoi, 
  citecolor=bluemoi, 
  urlcolor=bluemoi 
}

\makeatletter
\renewcommand{\figurename}{\sf \textbf{Figure}}
\renewcommand{\thefigure}{\arabic{figure}}
\renewcommand{\fnum@figure}{\sf\textbf{\figurename}~\textbf{\thefigure}}
\renewcommand{\tablename}{\sf\textbf{Table}}
\renewcommand{\thetable}{\arabic{table}}
\renewcommand{\fnum@table}{\sf\textbf{\tablename}~\textbf{\thetable}}
\makeatother

\begin{document}

\title{Entropy as a measure of attractiveness and socioeconomic complexity\\ in Rio de Janeiro metropolitan area} 

\author{Maxime Lenormand}
\thanks{These authors contributed equally to this work.}
\affiliation{TETIS, Univ Montpellier, AgroParisTech, Cirad, CNRS, INRAE, Montpellier, France}

\author{Horacio Samaniego}
\thanks{These authors contributed equally to this work.}
\affiliation{Laboratorio de Ecoinform\'atica, Universidad Austral de Chile, Instituto de Conservación, Biodiersidad y Territorio, Campus Isla Teja s/n, Valdivia, Chile}
\affiliation{Instituto de Ecolog\'ia y Biodiversidad, Facultad de Ciencias, Universidad de Chile, Las Palmeras, \~Nu\~noa, Santiago, Chile}
\affiliation{Instituto de Sistemas Complejos de Valpara\'iso, Subida Artiller\'ia 470, Valpara\'iso, Chile}

\author{Julio C. Chaves}
\affiliation{Getulio Vargas Foundation, Praia de Botafogo 190, Botafogo, Rio de Janeiro, Brazil}

\author{Vinicius F. Vieira}
\affiliation{Federal University of S\~ao Joao Del Rey}

\author{Moacyr A. H. B. da Silva}
\affiliation{Getulio Vargas Foundation, Praia de Botafogo 190, Botafogo, Rio de Janeiro, Brazil}

\author{Alexandre G. Evsukoff}
\thanks{Corresponding authors: alexandre.evsukoff@coc.ufrj.br.}
\affiliation{Coppe/Federal University of Rio de Janeiro P.O. Box 68506, Rio de Janeiro, Brazil}

\begin{abstract}
Defining and measuring spatial inequalities across the urban environment remains a complex and elusive task that has been facilitated by the increasing availability of large geolocated databases. In this study, we rely on a mobile phone dataset and an entropy-based metric to measure the attractiveness of a location in the Rio de Janeiro Metropolitan Area (Brazil) as the diversity of visitors' location of residence. The results show that the attractiveness of a given location measured by entropy is an important descriptor of the socioeconomic status of the location, and can thus be used as a proxy for complex socioeconomic indicators.
\end{abstract}

\maketitle

\section*{Introduction}

While cities have long been recognized as the cradle of modern civilization by providing a safe place for cultural development, the inequality distribution of wealth and services remain the main pressing issue threatening the sustainability of modern societies. Despite the large technological advances making our life apparently easier, economic inequality has been on the rise worldwide since 1980. This has become such an issue that most recent datasets show that the top 1\% of the wealthy population capture twice as much of the global income growth as the bottom 50\% \cite{Alvaredo2018}. While such distribution disparity among urbanites and social stratification is currently under deep scrutiny among economists, including the spatial components to such descriptions, it imposes additional methodological difficulties given the vagility of human nature and the heterogeneity of the spatial distribution of resources.

While different views exist regarding the origins of socio-spatial inequalities across cities \cite{Ruiz2013}, the consequences of poorly integrated societies deeply affect opportunities in key realms of social life that hamper social cohesion at a local and societal levels \cite{Jargowsky1997, Massey1990, Wilson2012}. While some discuss causal factors behind socio-spatial inequalities, evidence coming from natural experiments have shown direct impacts on particularly vulnerable groups \cite{Cutler1997}. Such evidence, among others, has tied inequalities to societal imbalances leading to critical states in terms of security, health, and wealth distribution \cite{Ruiz2013, Cutler1997, Garreton2016, Krieger1999, Massey1988} dreading social cohesion and precluding possibilities of enriching the social capital at particular locations \cite{Bolt1998, Farber2015, Farber2013, Forrest2001}. Defining and measuring spatial inequalities remains a complex and elusive task for which scientists have recognized several dimensions that are, so far, poorly integrated with a general conceptual framework \cite{Louf2016, Massey1990, Netto2018}. For instance, its precise understanding is often linked to the study objects at hand and the particular methodology employed to study them. Dimensions of inequalities often include the localized concentration of particular groups within cities, the spatial homogeneity of social groups, their accessibility, or more particularly, their distance to downtown \cite{Caldeira2012}. Hence, devising appropriate tools to characterize the spatial distribution of complex socioeconomic factors may contribute to the urgently needed development of integrative urban planning.

The explosive use of Information and Communication Technologies (ICT), such as cellphones and large databases of user spending behavior, has made huge volumes of non-conventional data available for urban research purposes \cite{Batty2013, Bettencourt2014, Blondel2015, Louail2017, Barbosa2018}. Knowing the cellphone tower to which we connect permits the reconstruction of our daily trajectories, providing a surprisingly high spatio-temporal resolution of our social interactions \cite{Onnela2007, Panigutti2017}. This approach has been widely used recently to assess a variety of topics going from individual mobility patterns \cite{Gonzalez2008} and land use patterns \cite{Lenormand2015}, to the detection of relevant places of high social activity within the city \cite{Beiro2018}, thereby unveiling the structure and function of cities \cite{Louail2014, Lenormand2015, Sotomayor2020}. Devising an efficient mobility infrastructure has long been known as a means for city integration and the increasing availability of ICT data allows for a new understanding of spatial integration patterns and its relationship to mobility, socioeconomic and ethnic stratification \cite{Lamanna2018}. Such highly resolved datasets provide a contextual understanding of land use that is readily available to derive new measures of social integration in its spatial context, thereby contributing to accurate, and near-real-time, descriptions of urban dynamics \cite{Dannemann2018, Jiang2017, Motte2016, Rubim2013, Toole2015, Song2010}. Many of these studies are based on the concept of activity space \cite{Hagerstrand1970, Jiang2017, Schonfelder2003}, defined as the set of locations visited by a traveler throughout their daily activities. Different measures describing the activity space have been studied  to understand daily mobility patterns \cite{Phithakkitnukoon2012, Barbosa2018}. Among these metrics, metrics based on the Shannon entropy are particularly interesting to study human mobility patterns. Indeed, the concept of \enquote{Mobility Entropy} indicators has been widely used to measure the diversity of users' movement pattern \cite{Lin2012,Pappalardo2016,Vanhoof2018,Cottineau2019}. It can be used at different scales to evaluate the diversity of trips made by an individual \cite{Pappalardo2015,Pappalardo2016}, the diversity of locations visited by an individual \cite{Lin2012,Cottineau2019} or a group of individuals \cite{Lamanna2018,Lenormand2018}.

In this work, we rely on the concept of \enquote{Mobility Entropy} from the point of view of visiting locations in order to deepen our understanding of human mobility in the context of urban computing by focusing on the concept of attractiveness. We particularly look into mapping the entropy of urban structure using increasingly available mobile phone datasets as a tool to provide highly resolved descriptions of the relationship between attractiveness and several key aspects of the urban environment such as productivity, education and ethnic origin in the Rio de Janeiro Metropolitan Area of Brazil. We focus here on the diversity of visitors' residence to measure the attractiveness of a location and then compare our results to economic and social indicators to assess how entropy effectively relates to socioeconomic indicators. We show that entropy is an important descriptor of socioeconomic complexity across this vastly populated area. 

\begin{figure*}
	\centering
	\includegraphics[width=14cm]{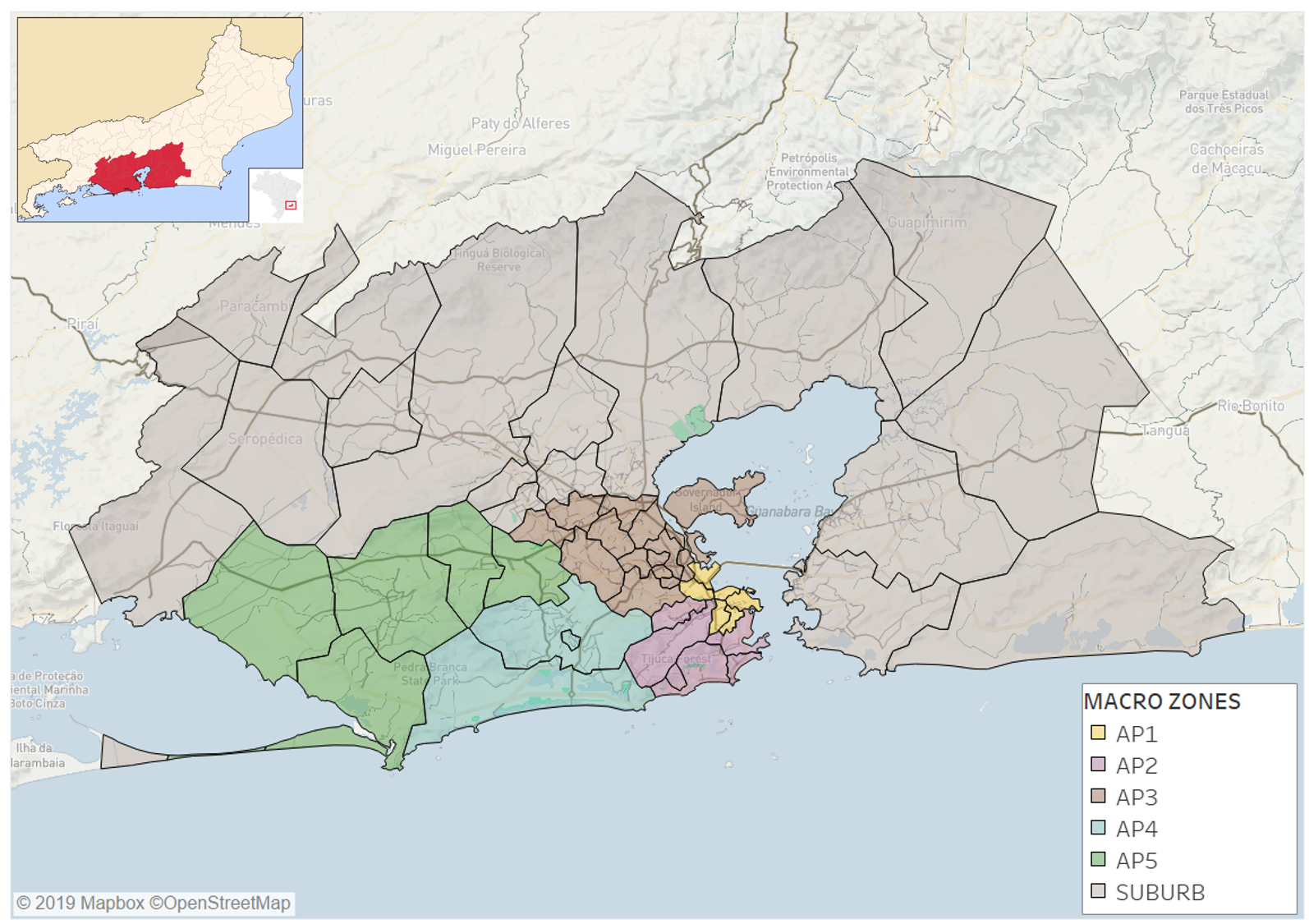}
	\caption{\textbf{Rio de Janeiro Metropolitan Area (RJMA).} The RJMA is composed of 49 locations, 16 municipalities outside the Capital represented in grey and 33 sub-districts inside the Capital, grouped into 5 districts.}
	\label{RJMA}
\end{figure*} 

\section*{Materials and Methods}

\subsection*{The study area and dataset}

The study area is the Rio de Janeiro Metropolitan Area (RJMA), the second largest urban area in Brazil with 12,145,734 inhabitants. Administratively, the RJMA is a part of the Rio de Janeiro State, of which Rio de Janeiro city (Rio for short) is the state Capital and the largest municipality with 6,320,446 inhabitants and 1,200,177 km$^2$. 

The organization responsible for the demographic census in Brazil is the Institute of Geography and Statistics (IBGE) who follows global standards to aggregate census tracts in sub-district, district, city, state, and country levels such that this partitioning can be used for most regions in the world and at different scales. This study relies on such partitioning, dividing the study area in 49 locations (Figure \ref{RJMA}) where the city of Rio is divided in 33 sub-districts aggregated into 5 districts as shown in Figure \ref{RJMA}. Districts are called Planning Areas (AP) and represent macro zones of the city with AP1 the center; AP2, the southern zone; AP3, the northern zone; AP4  Barra-Jacarepagu{\'a}; and AP5 depicts the western zone.

Our analysis is based on mobile phone data provided by a Brazilian telecommunication operator. The dataset was collected during 363 days between January and December 2014 across the phone area code 21. We use 2.1 $\times$ 10$^9$ call records originated from 2.9 $\times$ 10$^6$ anonymized subscribers. Only outgoing voice call data were made available for this work. We first focused on the identification of user's residence. The algorithm to detect places of residence is based on the analysis of the most frequently visited locations on evenings and weekends (see the Appendix for more details). This step allows us to discard users not living in the RJMA and remove users with no significant activity for the analysis. $350,685$ residences were identified.

We then aggregate the data in space and time. Aggregated records represent the number of users $v_{ij}(t)$ living in the location $i \in |[1,N]|$ and visiting the location $j \in |[1,N]|$ at time $t$. We spatially aggregate the antennas' Voronoi polygons in order to obtain $N=49$ locations matching the 49 locations composing the RJMA shown in Figure \ref{RJMA}. We also divide each day in four 6-hours shifts (Morning, Work, Afternoon and Night) and label each time period $t \in |[1,1452]|$ as either weekday or  weekend, including holidays. More details regarding the data preprocessing are available in Appendix.

\subsection*{Entropy as a measure of attractiveness} 
\label{entropy}

For each time interval $t$, there is a probability that a user living in $i$, will visit location $j$ described by:

\begin{equation}
p_{i \rightarrow j}(t)=\frac{v_{ij}(t)}{\sum_{k=1}^N v_{ik}(t)}
\label{pij}
\end{equation}

This probability describes the production of visitors and is normalized by the total number of users living in location $i$. In this study, we are interested in the diversity of visitors' location of residence as a measure of the attractiveness of the destination. We therefore need to compute the probability $p_{j \leftarrow i}$ that a user visiting location $j$ lives in location $i$. To do so, we combine the probability $p_{i \rightarrow j}$ with census data to estimate $V_{ij}(t)$, the number of users living at location $i$ and visiting the location $j$ at time $t$ using the following equation.

\begin{equation}
V_{ij}(t)=O_i p_{i \rightarrow j}(t),
\label{Vij}
\end{equation}

$O_i$ is the population of location $i$ according to the 2010 IBGE census. We can now compute the probability $p_{j \leftarrow i}(t)$ for an individual visiting $j$ at time $t$ that lives in $i$ (Equation \ref{pji}). 

\begin{equation}
p_{j \leftarrow i}(t)=\frac{V_{ij}(t)}{\sum_{k=1}^N V_{kj}(t)}
\label{pji}
\end{equation}

This second probability is thus related to the attraction of visitors, normalized at destination, and allows us to compute the normalized Shannon entropy as follows, 

\begin{equation}
S_j(t)=\frac{-1}{log(N)}\sum_{k=1}^N p_{j \leftarrow k}(t) log(p_{j \leftarrow k}(t))
\label{Sj}
\end{equation}

Large entropy values ($S_j(t) \approx 1$) mean that people visiting location $j$ at time $t$ are evenly distributed among all 49 locations, whereas a smaller values of entropy means that people visiting location $j$ at time $t$ tend to be mostly concentrated among few residence locations. The entropy has been widely used to analyze and model human mobility patterns. It can be used in spatial analysis to describe the diversity of individual movement patterns \cite{Vanhoof2018} or in spatial interaction modeling to estimate trip distributions by entropy maximization \cite{Wilson1969} to name a few. It is worth noting that we focus in this work on the analysis of entropy as a measure of attractiveness that can be used as a proxy for complex socioeconomic indicators.

It is important to keep in mind that a given entropy value can cover a large variety of situations regarding the distance traveled by visitors. Here, we characterize the relationship between traveled distance and entropy by computing the radius of attraction of a location $j$ as the average distance traveled by people visiting $j$ at time $t$:

\begin{equation}
R_j(t)=\sum_{k=1}^N p_{j \leftarrow k}(t) d_{kj},
\label{Dj}
\end{equation}

where $d_{kj}$ is the distance from location $k$ to $j$ along the road network between the locations' centroids computed using the Google Maps API \footnote{ \url{https://developers.google.com/maps/documentation/distance-matrix/}}. This calculation is particularly important in the case of Rio due to the presence of mountains, lakes and the Guanabara Bay, which makes road distances between certain locations very different from the Euclidean distances.

Finally, we also consider the ratio between the number of visitors divided by the population as a complementary measure of attractiveness.

\begin{equation}
\delta_j(t) = \frac{ \sum_{k=1}^N V_{kj}(t) }{ O_j }
\label{Deltaj}
\end{equation}

\begin{figure*}
	\centering
	\includegraphics[width=14cm]{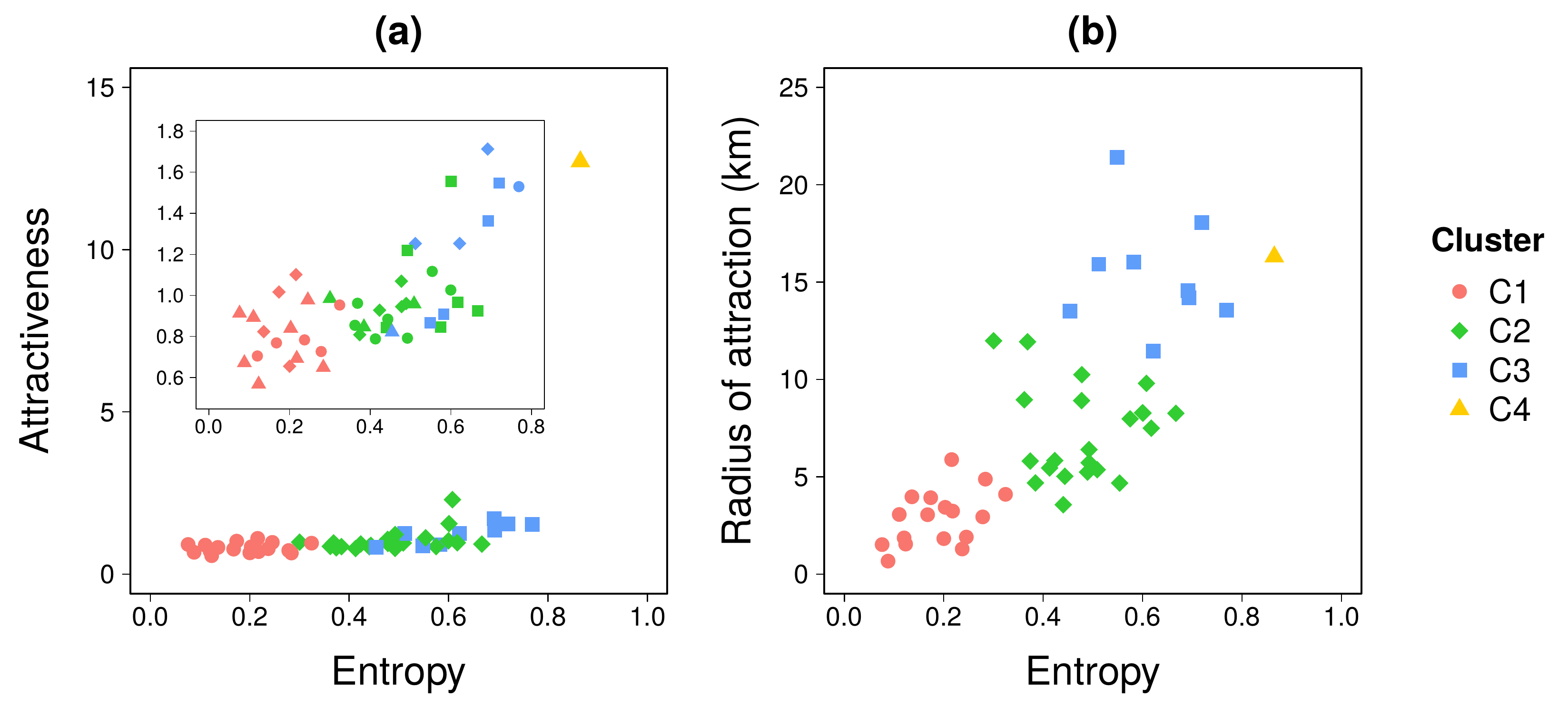}
	\caption{\textbf{Results of the clustering analysis.} Log-log scatter plot of \textbf{(a)} the attractiveness and \textbf{(b)} the radius of attraction in terms of the entropy index. The inset in (a) shows the relationship after removing one outlier (cluster C4). Each dot represents a location within the study area. Indicators have been averaged over the work shift time period during weekdays.}
	\label{Clustering}
\end{figure*} 

\subsection*{Entropy, economic and sociodemographic indicators}

Because our entropy index represents a synoptic representation of mobility across the RJMA, we finally seek to describe its impact in terms of well known economic, social and demographic indicators as collected by the IBGE. We therefore evaluated how the diversity of visitors relates to the economic performance of the city by plotting the number of jobs and income levels against our entropy estimation. Sociodemography, in turn, was assessed by establishing the relationship between education levels in primary and secondary (high school) education among the population resident in each partitioned area. Finally, two developmental indices were chosen to evaluate entropy performance across the RJMA. 


\section*{Results}

\subsection*{Classification of locations according to their attractiveness}

We start our analysis by performing a clustering analysis to group together locations exhibiting similar features regarding their attractiveness. As a first step, we focus on two features across the urban landscape, the diversity at the origin location and the attractiveness at work locations. This led us to average the three indicators for each location (Equations \ref{Sj}, \ref{Dj} and \ref{Deltaj}) over the work shifts time periods on weekdays. Locations are clustered using the k-means algorithm based on the three standardized averaged metrics. The number of clusters were chosen based on the ratio between within-group variance and the total variance (see the Appendix for more details). We obtained four clusters. Clustering results and the relationships between the different metrics are shown in Figure \ref{Clustering}. We observe a positive relationship between metrics, in which attractiveness and radius of attraction tend to increase with the entropy. There is nevertheless a strong dispersion around these tendencies with an attractiveness and radius of attraction values that can double for a given entropy value.

\begin{figure*}[!ht]
	\centering
	\includegraphics[width=13cm]{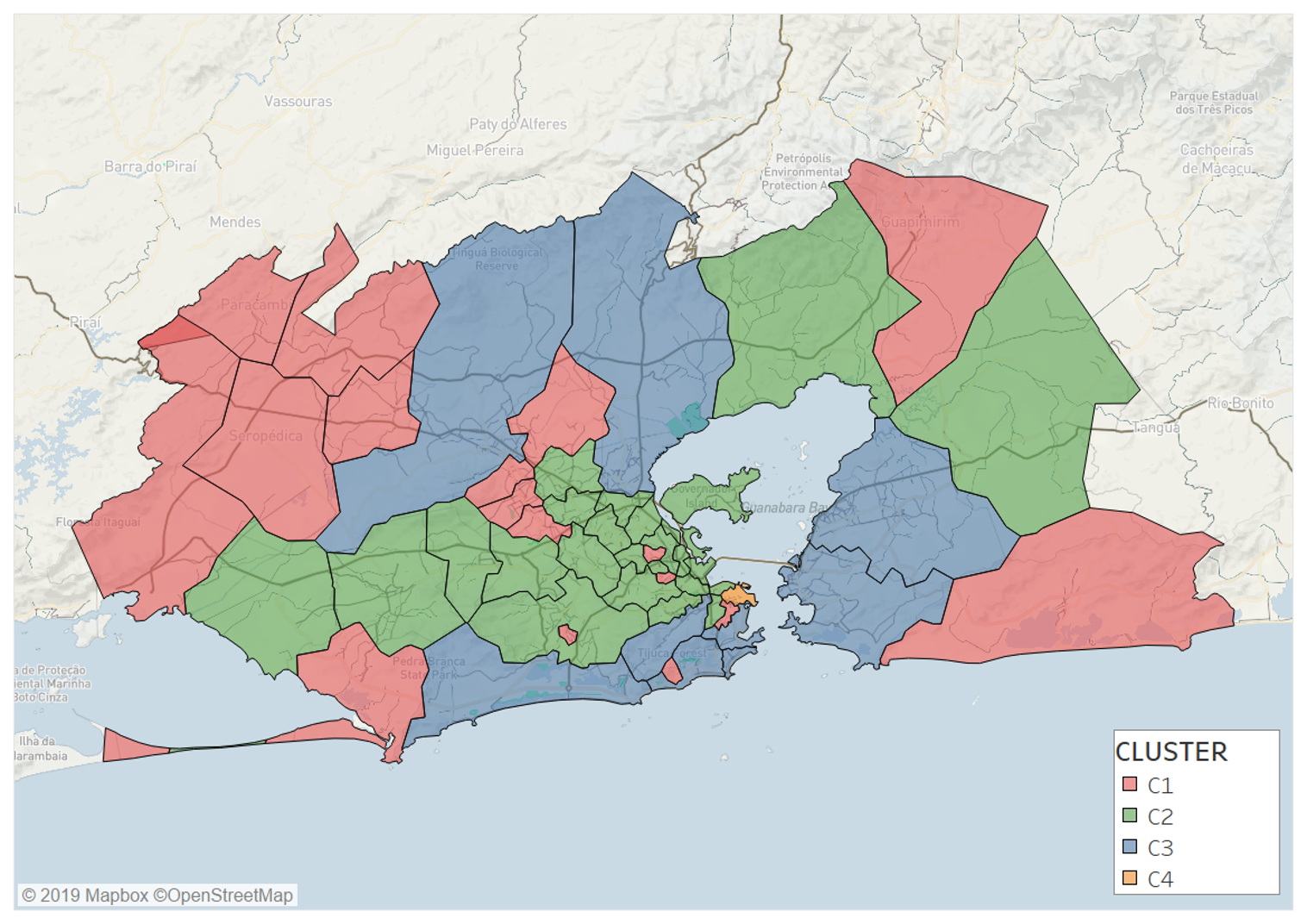}
	\caption{\textbf{Map of the RJMA that display the spatial distribution of four clusters.}}
	\label{MapClust}
\end{figure*}

Figure \ref{MapClust} shows the spatial distribution of the four resulting clusters across the whole studied area. Clusters are determined by a certain level of attractiveness and can be described as follows: 

\begin{itemize}
	
	\item \textbf{C1 (red)} represents a low attractive cluster composed of 17 locations. It is characterized by a low entropy, an attractiveness ratio lower than one and a low radius of attraction. Locations in C1 are far from the Rio city center or segregated areas inside the Capital;
	
	\item \textbf{C2 (green)} is a cluster of 22 locations, mostly located inside the city. This cluster is characterized by medium values of entropy of visitors and radius of attraction, while having an attractiveness ratio close to one; 
	
	\item \textbf{C3 (blue)} is an attractive group with 8 locations mostly near to the sea inside Capital. This cluster shares high entropy values, attractiveness ratio between 1 and 2 and a large radius of attraction;
	
	\item \textbf{C4 (orange)} is composed of only one location that can be considered as an outlier due to its very high attractiveness. The remaining three clusters do not change if this outlier is removed before clustering. This location is the business center (Centro) of the city, and is a very attractive cluster with a very large entropy ($S_{C4} \approx 0.9$), attractiveness ratio and radius of attraction ($\delta_{C4} \approx 12$). This location concentrates most of jobs and visitors from all the RJMA. 
	
\end{itemize}

\begin{figure*}[!ht]
	\centering
	\includegraphics[width=14cm]{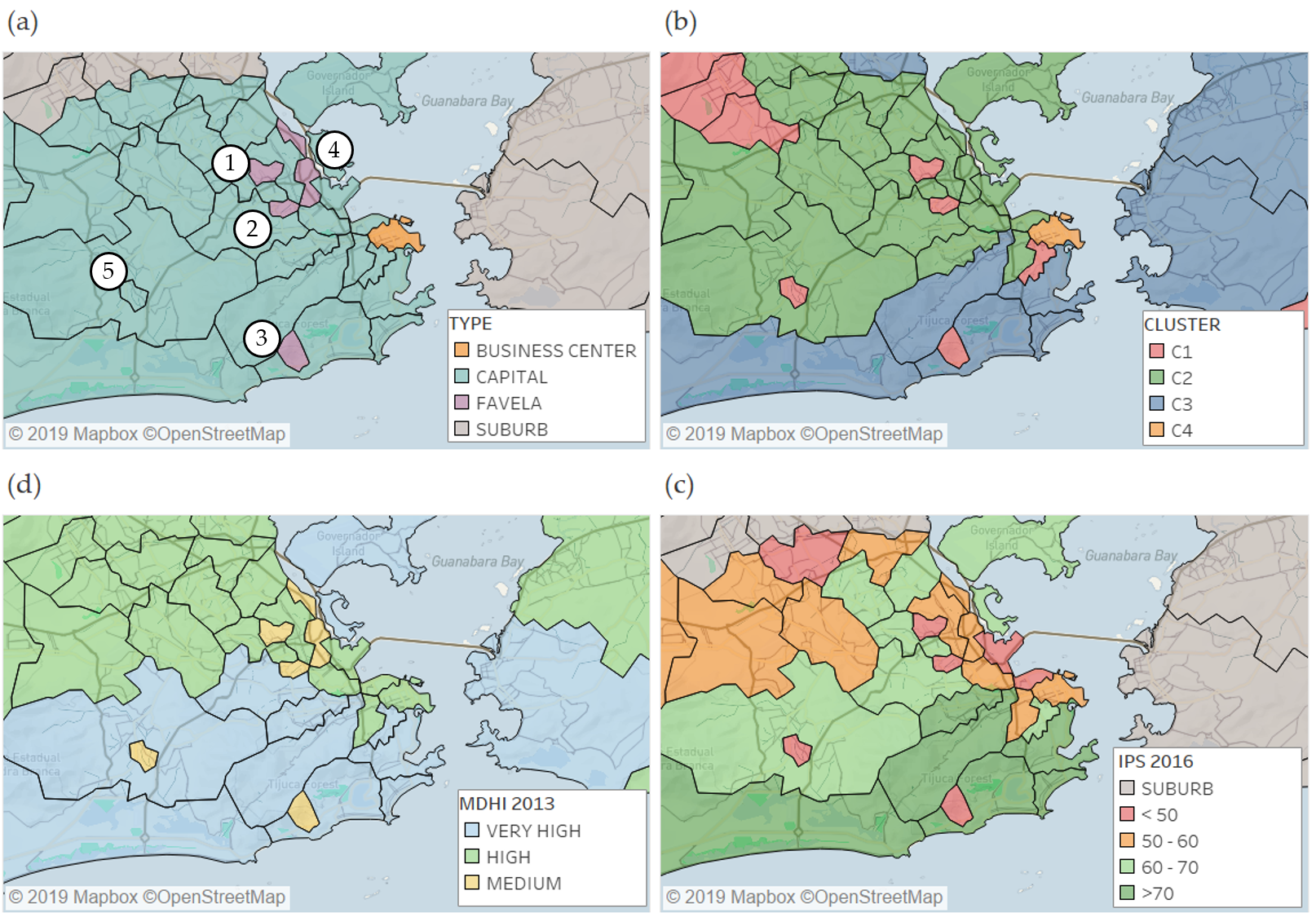}
	\caption{\textbf{Zoom on Rio de Janeiro city.}  (a) Favela sub-districts (in purple) and business center (in orange) locations. We focus the discussion in five locations (1) Complexo do Alemão, (2) Jacarezinho, (3) Rocinha, (4) Complexo da Maré, and (5) Cidade de Deus. (b) Clusters spatial distribution. (c)  Municipal Human Development Index (MHDI) from 2013. (d) Social Progress Index (IPS) from 2016.} 
	\label{MapClust_ZOOM}
\end{figure*} 

Our methodology allows us to detect segregated areas with a very low diversity of visitors and attractiveness. Figure \ref{MapClust_ZOOM} shows the comparison of the clustering results with two social development indexes.  We focus the discussion in five locations shown in Figure \ref{MapClust_ZOOM}a: (1) Complexo do Alemão, (2) Jacarezinho (3) Rocinha (4) Complexo da Mar{\'e} and (5) Cidade de Deus. The first four locations are classified by Rio City Hall as favela sub-districts\footnote{\url{http://bit.ly/2O9SEdA} (in portuguese)} and are shown in purple in Figure \ref{MapClust_ZOOM}a. The term \enquote{favela} is used here in the sense of subnormal agglomerate as defined by IBGE \footnote{\url{http://bit.ly/337gQlb}}: \enquote{a form of irregular occupation of land usually characterized by an irregular urban pattern, with scarce essential public services and located in areas not proper or allowed for housing use}. In a broad sense, favela also includes urbanized areas, areas that were once subnormal agglomerates but have been urbanised, and also housing estates. The favela sub-districts assigned in purple in Figure \ref{MapClust_ZOOM}a are defined according to Rio City Hall, as the locations with more that 50\% of population living in subnormal agglomerates. In Cidade de Deus, only 13\% of the population is living in subnormal agglomerates as it is mostly composed by housing estates building, while its socioeconomic indexes are similar to the favela sub-districts.

\begin{figure*}[!ht]
	\centering
	\includegraphics[width=13cm]{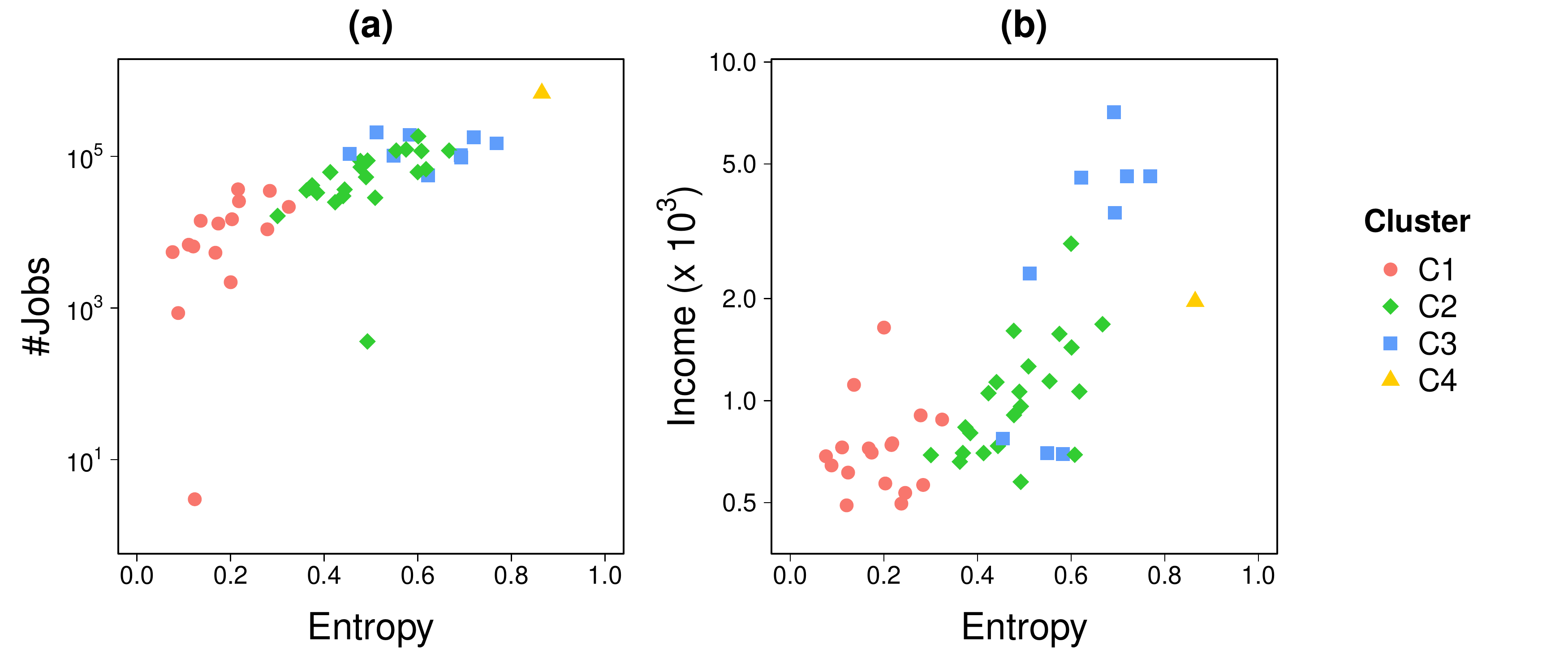}
	\caption{\textbf{Economic analysis.} Number of jobs (a) and income (in Brazilian Reals) (b) as a function of the entropy index. The entropy have been averaged over the work shift time periods on weekdays.}
	\label{Socioeco1}
\end{figure*} 

Figure \ref{MapClust_ZOOM}b shows a zoom in the clustering results. The main favela sub-districts were classified in low attractive cluster (C1) as did Cidade de Deus. Complexo da Mar{\'e} has also many housing estates building and 54\% of its population living in subnormal agglomerate. It was classified in the medium attractive cluster (C2), maybe because it is crossed by two of the main expressways that lead to the exit of the city. In the dataset used in this work, a visitor is detected in a given location by a call recorded within the location, such that some detected visitors may be passing-by the location to reach another destination.

Figure \ref{MapClust_ZOOM}c and \ref{MapClust_ZOOM}d show two social development indexes. In Figure \ref{MapClust_ZOOM}c the Municipal Human Development Index (MHDI), which is an adaptation of the Human Development Index (HDI) for municipalities. The MHDI data were obtained from the Atlas of Human Development in Brazil \footnote{\url{www.atlasbrasil.org.br}}, where the MHDI computed in 2013 is available at the census track level, as so as aggregated values for all municipalities and for district level in  metropolitan areas. In Rio, the MHDI is available for the macro zones shown in Figure \ref{RJMA} and the value for the five locations of interest in Figure \ref{MapClust_ZOOM}a were obtained from the census track level. The classes and colours used in Figure \ref{MapClust_ZOOM}c were suggested by the Atlas. All five locations assigned in Figure \ref{MapClust_ZOOM}a were classified as medium MHDI and many locations classified in the high attractive cluster (C3) have very high MHDI.

The MHDI is a global index intended to compare the social development in the whole country. The Rio City Hall has adopted the Social Progress Index (IPS), which is more focused on the city characteristics and is based in 32 indicators in three dimensions. The data used in this work were computed in 2016 and obtained from the open data portal of Rio City Hall \footnote{\url{www.data.rio}}. The colours and levels presented in Figure \ref{MapClust_ZOOM}d are the ones used by the Rio City Hall. It can be seen from Figure \ref{MapClust_ZOOM}d that all four locations assigned in low attractive cluster (C1) have low IPS (IPS $\leq$ 50). The Complexo da Mar{\'e} sub-district has medium IPS (50 $\leq$ IPS $\leq$ 60) and was assigned to the medium attractive cluster (C2). Moreover, most locations assigned to high attractive cluster (C3) have a very high IPS (IPS $\geq$ 70). There is a very good agreement between the clusters computed from mobility and IPS, as cluster C1 correspond to IPS $\leq$ 50, cluster C2 corresponds to 50 $\leq$ IPS $\leq$ 70 and cluster C3 corresponds to IPS $\geq$ 70. 

In the next section, we discuss the relationship between the mobility indicators and the economic and social indicators selected for this study.

\begin{figure*}[!ht]
	\centering
	\includegraphics[width=13cm]{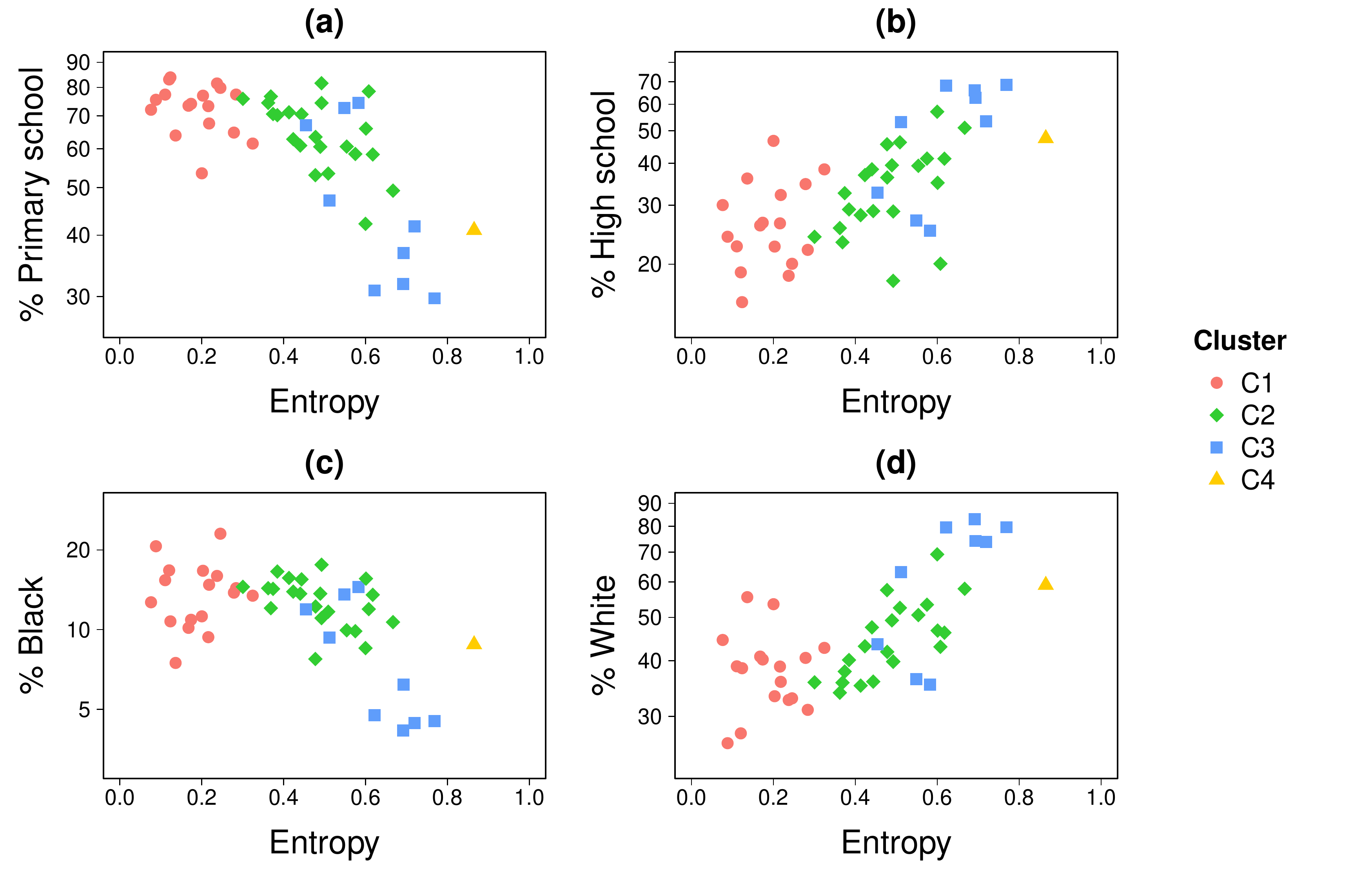}
	\caption{\textbf{Sociodemographic analysis.} Percentage of primary school level education (a), high school level education (b), black people (c), and white people (d) as a function of the entropy index. The entropy have been averaged over the work shift time periods on weekdays.}
	\label{Socioeco2}
\end{figure*} 

\begin{figure*}[!ht]
	\centering
	\includegraphics[width=13cm]{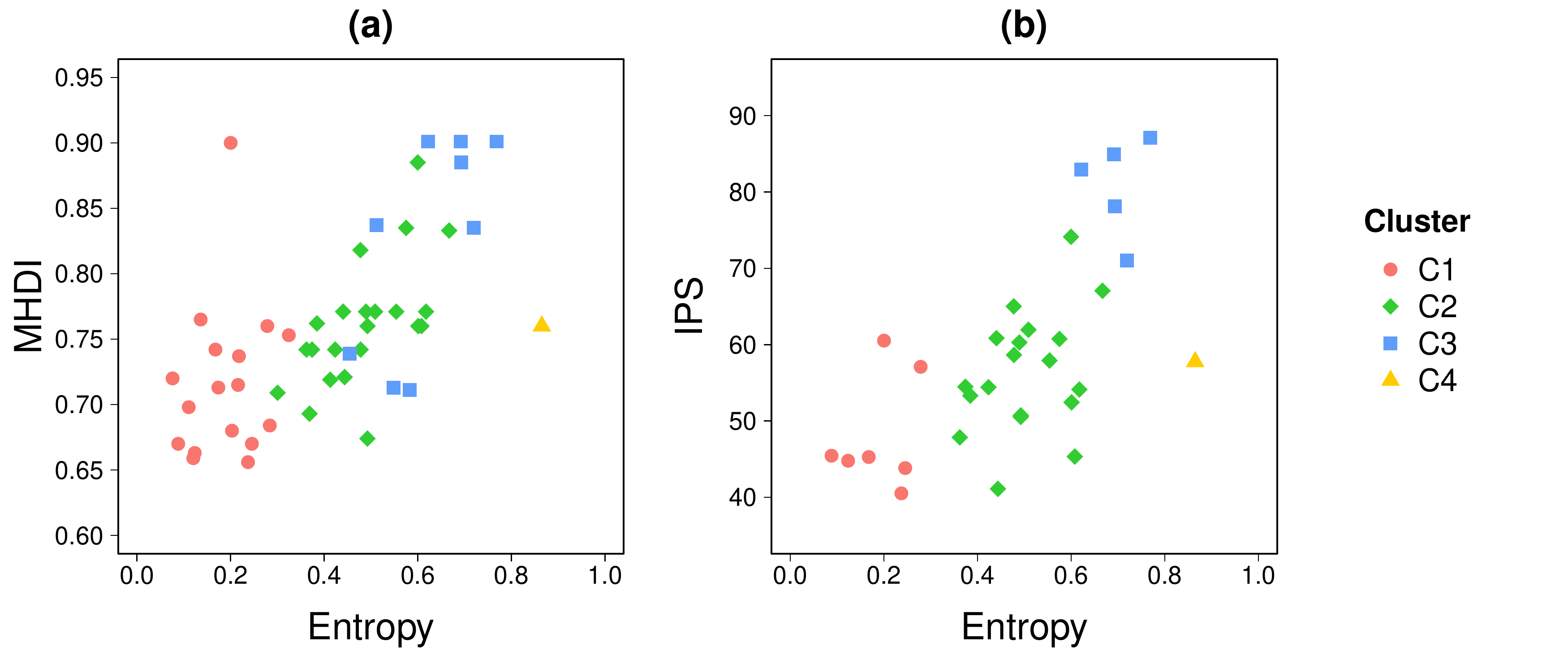}
	\caption{\textbf{Social development indexes.} MHDI (a) and IPS (b) as a function of the entropy index. The entropy have been averaged over the work shift time periods on weekdays.}
	\label{IDHM}
\end{figure*} 

\subsection*{Economic activity and sociodemographic factors}

While transportation mobility has largely been recognized as a major player in the urban economy \cite{Duranton2004}, the recent scrutiny of Call Detail Records (CDR) have expanded our understand of how mobility relates to economic activity across cities \cite{Cottineau2019,Xu2018}. We here evaluated how entropy relates to officially reported job numbers and income levels (Figure \ref{Socioeco1}). In spite of the large informal job market known to occur in RJMA, our analysis shows a positive and exponential relationship between formal jobs and entropy (Figure \ref{Socioeco1}a). Similar patterns emerge when relating income level with entropy (Figure \ref{Socioeco1}b) as well as with Gross Domestic Product (GDP) (see Appendix).

Interestingly, opposite trends emerge when entropy is plotted against demographic indices such as the percentage of the population having completed primary education and high school degrees. In Figure \ref{Socioeco2}, \enquote{primary school} refers to the percentage of individuals having primary school or lower education level and \enquote{high school} refers to individuals having high school or higher education level. School degrees are positively correlated with income, meaning that higher income locations tend to have higher education levels. In the same way race is negatively correlated with income, there is indeed a prevalence of white skin individuals in higher income locations and the prevalence of black skin individuals in lower income locations. As entropy is related to income (Figure \ref{Socioeco1}), locations having a large fraction of its population with a completed primary school diploma or lower exhibit lower entropy values (Figure \ref{Socioeco2}a), while locations with a large proportion with high school or higher education level is positively associated to entropy (Figure \ref{Socioeco2}b). This is strikingly similar to the pattern exhibited by ethnic origin. Black skin population, as well as the percentage of primary school, also shows a negative relation to entropy (Figure \ref{Socioeco2}c), while areas with a larger percentage of white skin population tend to exhibit higher entropy values (Figure \ref{Socioeco2}d).

The entropy of visitors, computed from CDR, reflects the complexity of indicators usually computed using classical approaches. In fact, entropy seems to be positively associated with socioeconomic indicators such as MHDI and IPS (Figure \ref{IDHM}), as shown in Figures \ref{Socioeco1} and \ref{Socioeco2}.

\begin{figure*}[!ht]
	\centering
	\includegraphics[width=12cm]{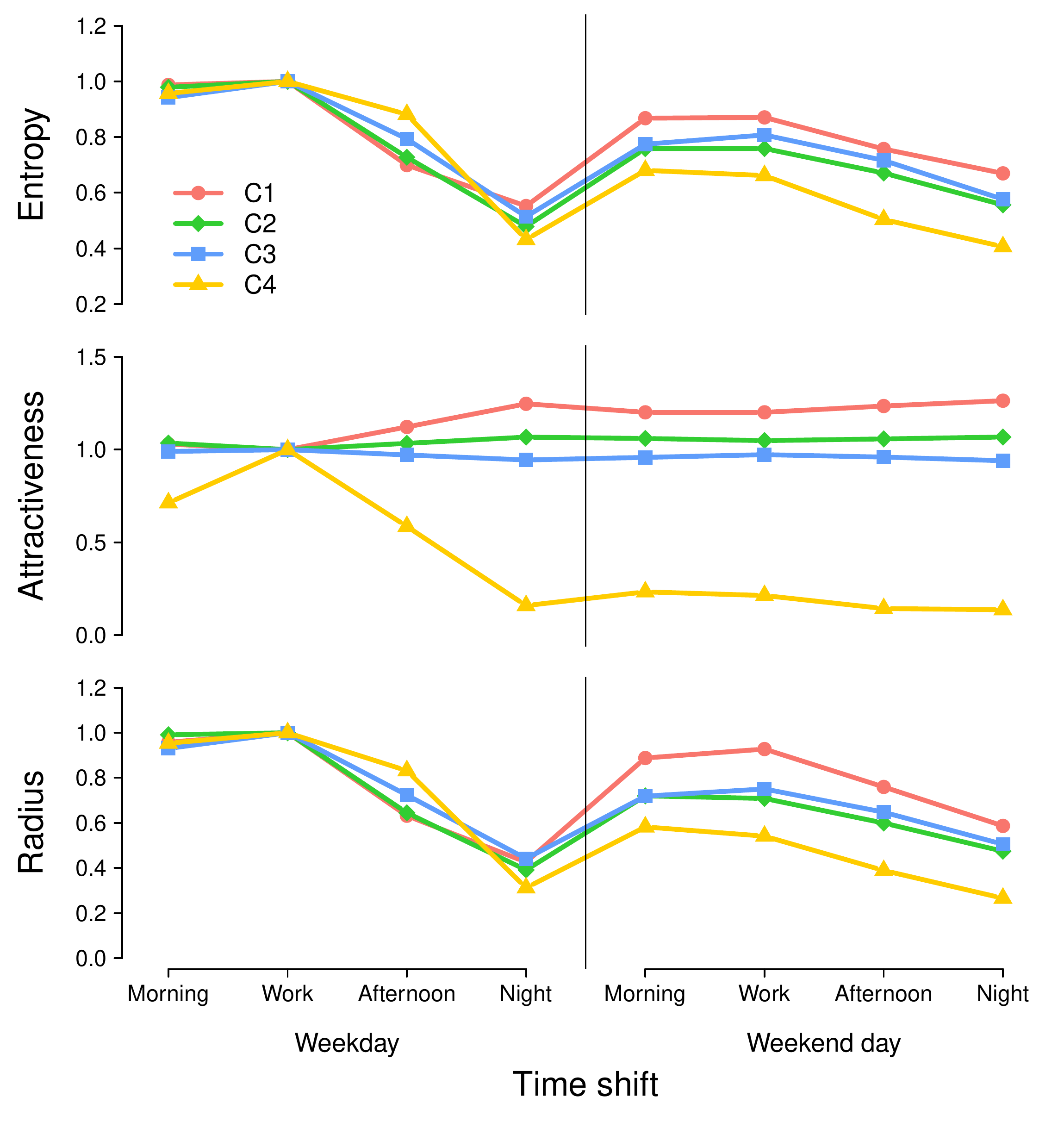}
	\caption{\textbf{Temporal evolution of the three metrics.} From the top to the bottom, entropy, attractiveness and radius of attraction as a function of time by cluster. The values are averaged by cluster and normalized by the value obtained for the work shift during weekdays. A similar plot displaying boxplots instead of average values is available in Appendix.}
	\label{Time}
\end{figure*} 

\subsection*{Temporal evolution of the attractiveness}

To study the temporal evolution of entropy, attractiveness and radius of attraction, we plot the normalized average metric values for each cluster across time shifts (Figure \ref{Time}). Normalizations are performed using the reference values obtained for the work shift time period on weekdays. We decided here to consider relative, instead of absolute, values in order to make average attractiveness of clusters of locations comparable over time. Entropy tends to globally decrease along the day on both weekdays and weekends for every location whatever the cluster it belongs to. It is, however, interesting to note that the entropy is relatively higher during weekday night and weekends for locations classified as low attractive during weekday work shifts compared to highly attractive locations. Indeed, while locations of cluster C4 exhibits an entropy index 50\% lower than the reference value, it actually represents 80\% for cluster C2/C3, and more than 90\% for locations belonging to cluster C1. A similar behavior is observed for the radius of attraction. The situation is slightly different, however, for the attractiveness with an increase of the metrics during afternoons and night shifts on weekdays for the low attractive cluster C1. It further reaches a plateau during the weekend days. The location of cluster C4 shows an opposite behavior with a decreasing attractiveness along the day to reach a plateau during weekend days. The attractiveness remains more or less constant for locations belonging to cluster C2 and C3.

\section*{Discussion}

The impact of socio-spatial inequalities on urban systems has largely been treated in the urban economics and sociological literature, but the increasing availability of large mobile phone databases has open the possibility to provide a clearer picture of how different aspects of urban life impact economic and sociodemographic aspects of cities \cite{Blondel2015}. Going into this direction, this work presents the results of the processing of 2.1 Billion records collected from 2 million users in the Rio de Janeiro Metropolitan Area, Brazil, during the whole year of 2014, placing this research among the largest analysis, to our knowledge, used to relate mobility and its link to socioeconomic complexity in Brazil. We hereby illustrate the potential of combining mobile phone data with entropy-based metrics to measure the attractiveness of a location. This may prove useful to urban planners and managers when it comes to describe and plan for complex socioeconomic indicators. While it is known that mobility is in fact related to economic activity, this work presents an effective and simple way to measure such relationships from increasingly available ICT data such as mobile phone datasets. 

While most capital cities in South America suffer from a disproportionate growth compared to other urban settlements \cite{Henderson1991}, common patterns of spatial inequalities show that underprivileged populations establish themselves away from highly productive central zones \cite{Sabatini2006,Dannemann2018}, with often clear differences among the usage of urban infrastructure \cite{Lotero2016}. In this sense, the particular and complex topography of Rio de Janeiro would suggest the existence of shared usage patterns of the city among urbanites coming from different social contexts. The spatial partitioning employed in our study closely matches IBGE delineation, we are therefore able to compare official statistics with measures derived from CDR data and offer specific insights regarding the usage of ICT as proxies for the spatial distribution of complex socioeconomic indicators derived from mobile phone datasets. Our analysis shows that the attractiveness of a district measured with the diversity of visitors' place of residence is correlated with the income and the number of jobs in spite of the large informal job market of Rio \cite{Motte2016}. 

We also show that the attractiveness is lower in areas hosting a large percentage of the population with African descent and/or locations where primary school training is prevalent (Figure \ref{Socioeco2}a,c). While this points to previous descriptions showing how available schooling options closely reproduce residential patterns of socio-spatial segregation \cite{Flores2008, Li2013}, the spatial mismatch and highly productive Centro area, where work opportunities are concentrated in the RJMA, leads us to think that residential segregation of the poorest is reinforced by new inequalities when taking into account daily mobility opportunities. Unfortunately, and in spite of using state-of-the-art descriptors of urban diversity, we are able to corroborate a well-known trend in which areas with large African descendant populations are still syndicated as an indicator of social inequality. This poses important planning challenges to historical areas such as the RJMA, where almost one million enslaved Africans were estimated to arrive in the XVII$^{th}$ century \cite{Karasch1987}. 

The observed results concur on recent developments in the scientific literature that show how mobile phone information can be used to evaluate the socioeconomic state of spatially heterogeneous regions \cite{Pappalardo2015,Eagle2010,Blumenstock2015}, especially in developing countries. Moreover, the RJMA is a very particular case study where socioeconomic isolated districts are placed in-between richer areas, as well as in the periphery, which is more common in greater cities of developing countries. This particular characteristic of the city allows to validate the results, as the clusters accurately identified favelas and other socioeconomic isolated districts, as shown in Fig. \ref{MapClust_ZOOM}.

In summary, this manuscript serves to illustrate the potential of mobile phone data combined with entropy-based metrics for measuring the attractiveness of a location that can be used as a proxy for complex socioeconomic indicators. Even if the spatial partitioning used in this study tends to reduce the level of spatial uncertainty inherent in this type of data sources \cite{Lenormand2016}, it would be interesting to reproduce the results with different datasets coming form different sources of mobility information.

\vspace*{0.5cm}
\section*{Acknowledgements} 

JCC, VFV, MAHBS and AGE acknowledge the funding granted by The Rio de Janeiro State Research Agency (FAPERJ) and by the Getulio Vargas Foundation. The work of ML was funded by the French National Research Agency (grant number ANR-17-CE03-0003). HS was funded by FONDECYT-CONICYT Chile (grant no. 1161280).

\bibliographystyle{unsrt}
\bibliography{Entropy_CDR_Rio}

\onecolumngrid
\vspace*{2cm}
\newpage
\onecolumngrid

\makeatletter
\renewcommand{\fnum@figure}{\sf\textbf{\figurename~\textbf{S}\textbf{\thefigure}}}
\renewcommand{\fnum@table}{\sf\textbf{\tablename~\textbf{S}\textbf{\thetable}}}
\makeatother

\setcounter{figure}{0}
\setcounter{table}{0}
\setcounter{equation}{0}

\section*{Appendix}

\subsection*{Data preprocessing}

\subsubsection*{Spatial aggregation}

\begin{figure}[h]
	\centering
	\includegraphics[width=12cm]{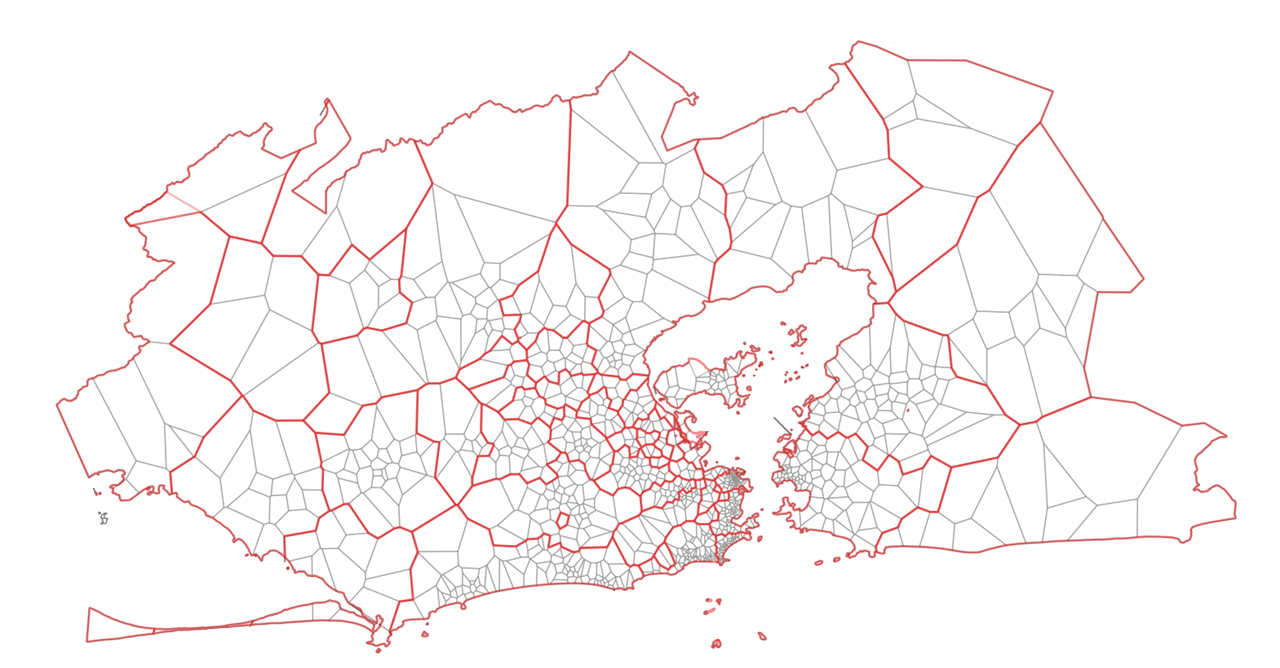}
	\caption{\textbf{Spatial distribution of antennas with their respective Voronoi polygons.}}
	\label{Voronoi}
\end{figure}

In this work, in order to link mobility results directly to socioeconomic data, each geographic unit is the union of Voronoi polygons of antennas (Figure A1) matching the geographic limits of the 49 locations (Figure S\ref{Voronoi}). This makes the set of regions outlined here directly related to the respective locations and, consequently, to the census data and many other sources.

\begin{figure}[H]
	\centering
	\includegraphics[width=12cm]{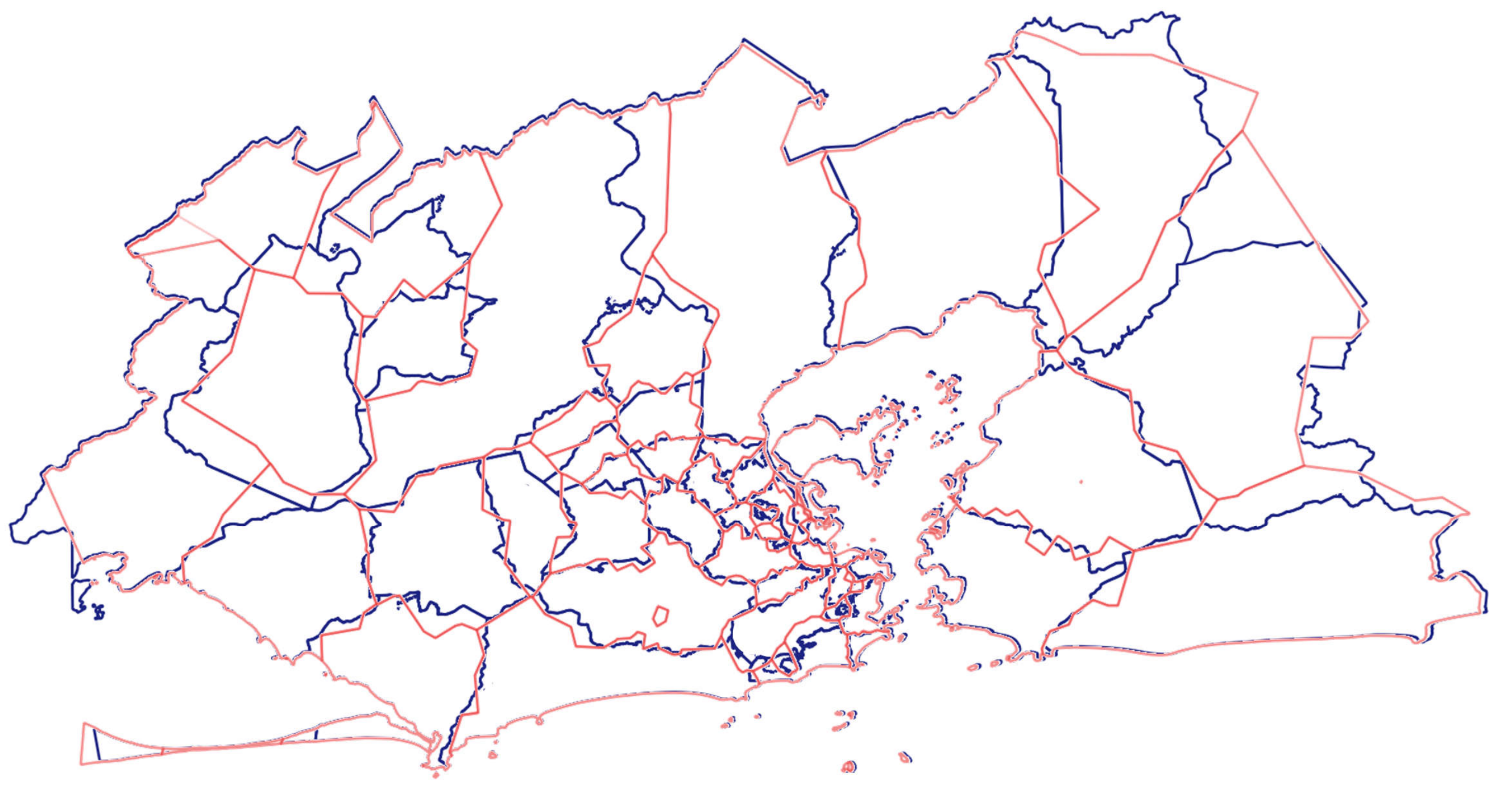}
	\caption{\textbf{Spatial overlap between the aggregation of Voronoi cells and the districts' spatial polygons.}}
	\label{Overlap}
\end{figure} 

\subsubsection*{Temporal aggregation}

\begin{figure}[H]
	\centering
	\includegraphics[width=12cm]{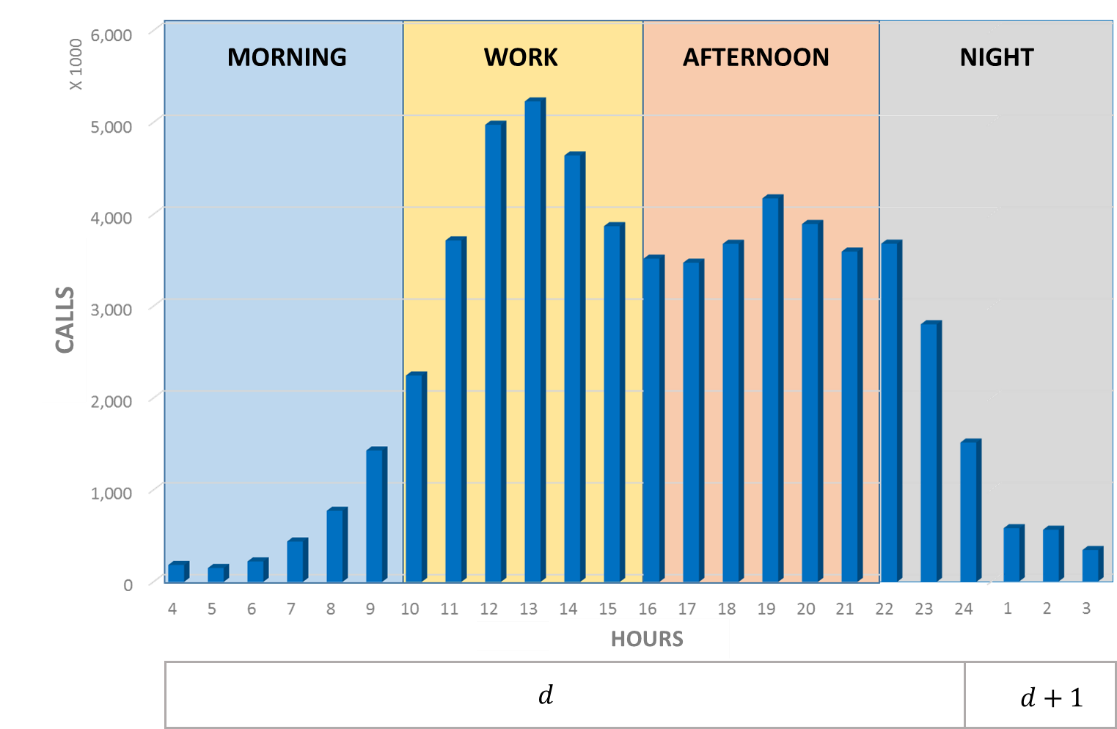}
	\caption{\textbf{Number of calls per hour and the partition of time shifts.} Total number of calls made in the RJMA in 2014 (including weekdays and weekends).}
	\label{Shift}
\end{figure}

In addition to the spatial partitioning, the results were aggregated for each day in the data set. The four shifts considered the distribution of activities throughout one day. The time shift with the smallest number of records was 04:00 AM, as shown in Figure S\ref{Shift}.

\subsubsection*{Identification of the user’s place of residence}

\begin{figure}[H]
	\centering
	\includegraphics[width=13cm]{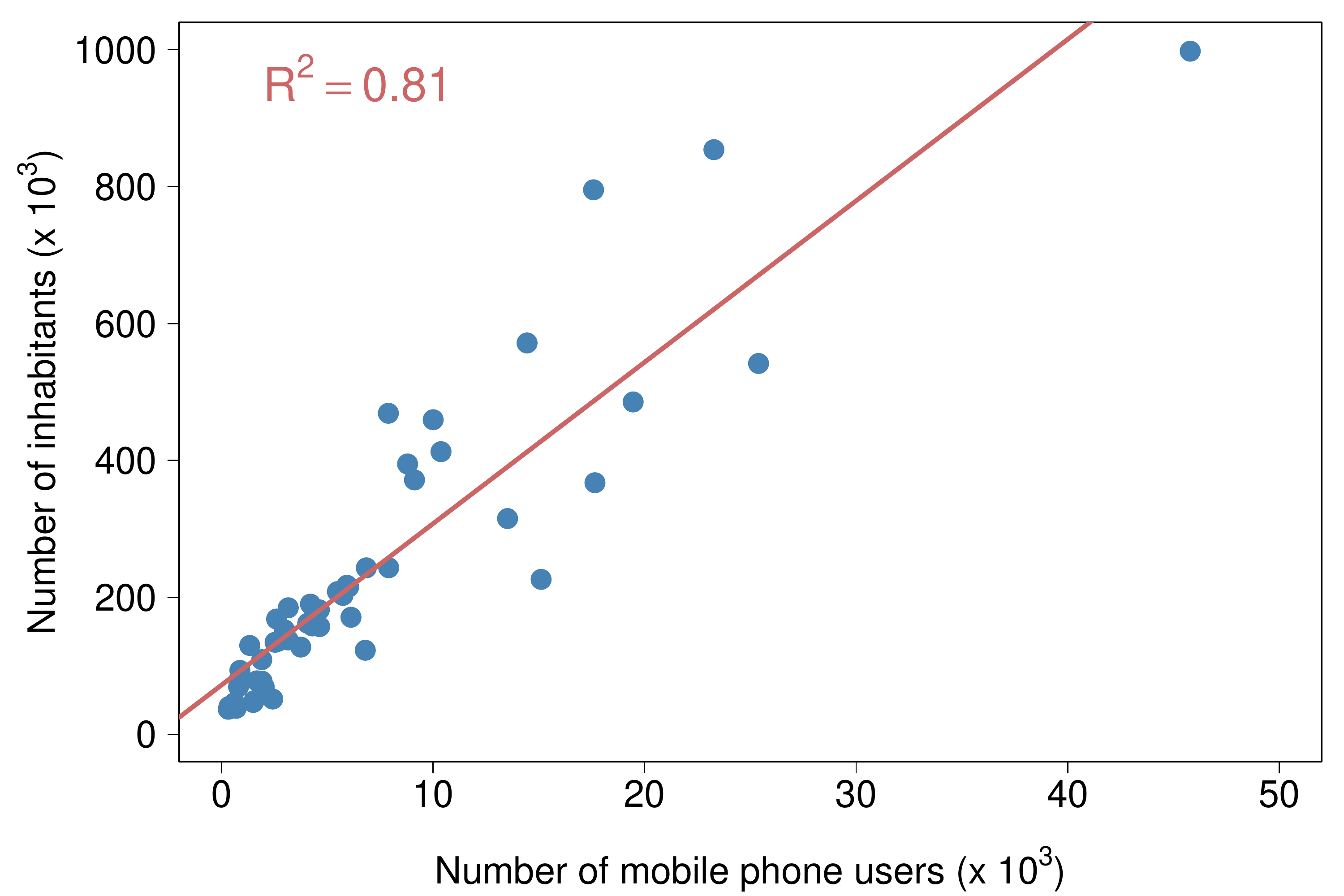}
	\caption{\textbf{Number of mobile phone users with an identified residence in the RJMA as a function of the number of inhabitants in the 49 locations.}}
	\label{Residence}
\end{figure}

The presumed residence of each user was computed as the most visited Voronoi cell between 08:00PM and 06:00AM during workdays and the entire day on Sundays and holidays. We additionally required that the user to be regularly detected in this cell (at least five times) and that the number of visits at the most frequented cell is always greater than the number of visits at the second most frequented cell. The final dataset containing only users with an identified residence ended up to be $350,685$ mobile phone users. As mentioned above, the data were aggregated spatially by assigning each Voronoi cell to one of the 49 districts. The identification of the user’s place of residence was then evaluated using data from the IBGE 2010 census. As it can be observed in Figure S\ref{Residence} we obtained a good match between the census data and the residence identified with mobile phone data with a Pearson correlation coefficient equal to 0.9.

\subsection*{Clustering analysis}

\begin{figure}[H]
	\centering
	\includegraphics[width=13cm]{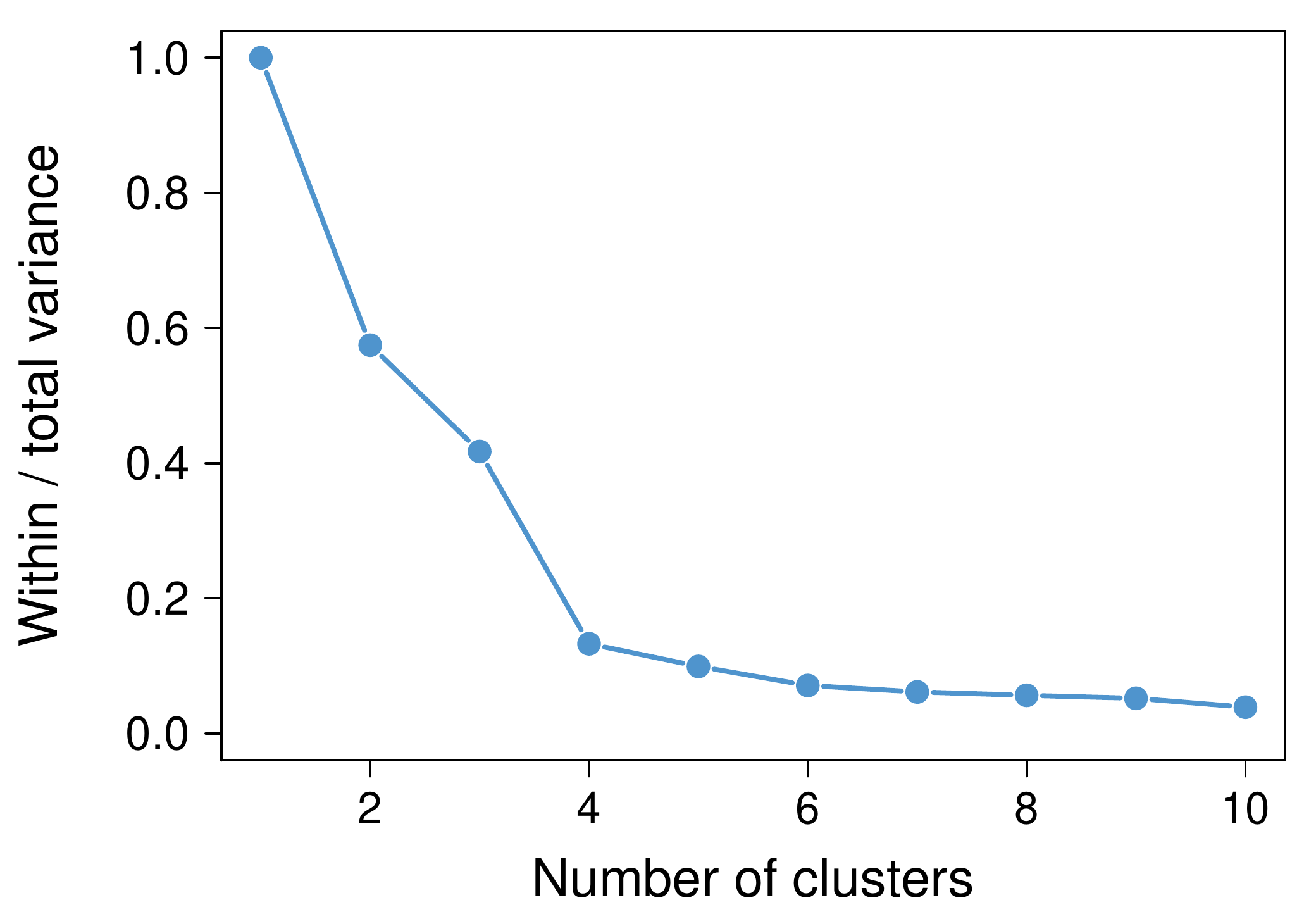}
	\caption{\textbf{Ratio between the within-group variance and the total variance as a function of the number of clusters.}}
	\label{Vintratot}
\end{figure} 

\subsection*{Economic activity}

\begin{figure}[H]
	\centering
	\includegraphics[width=13cm]{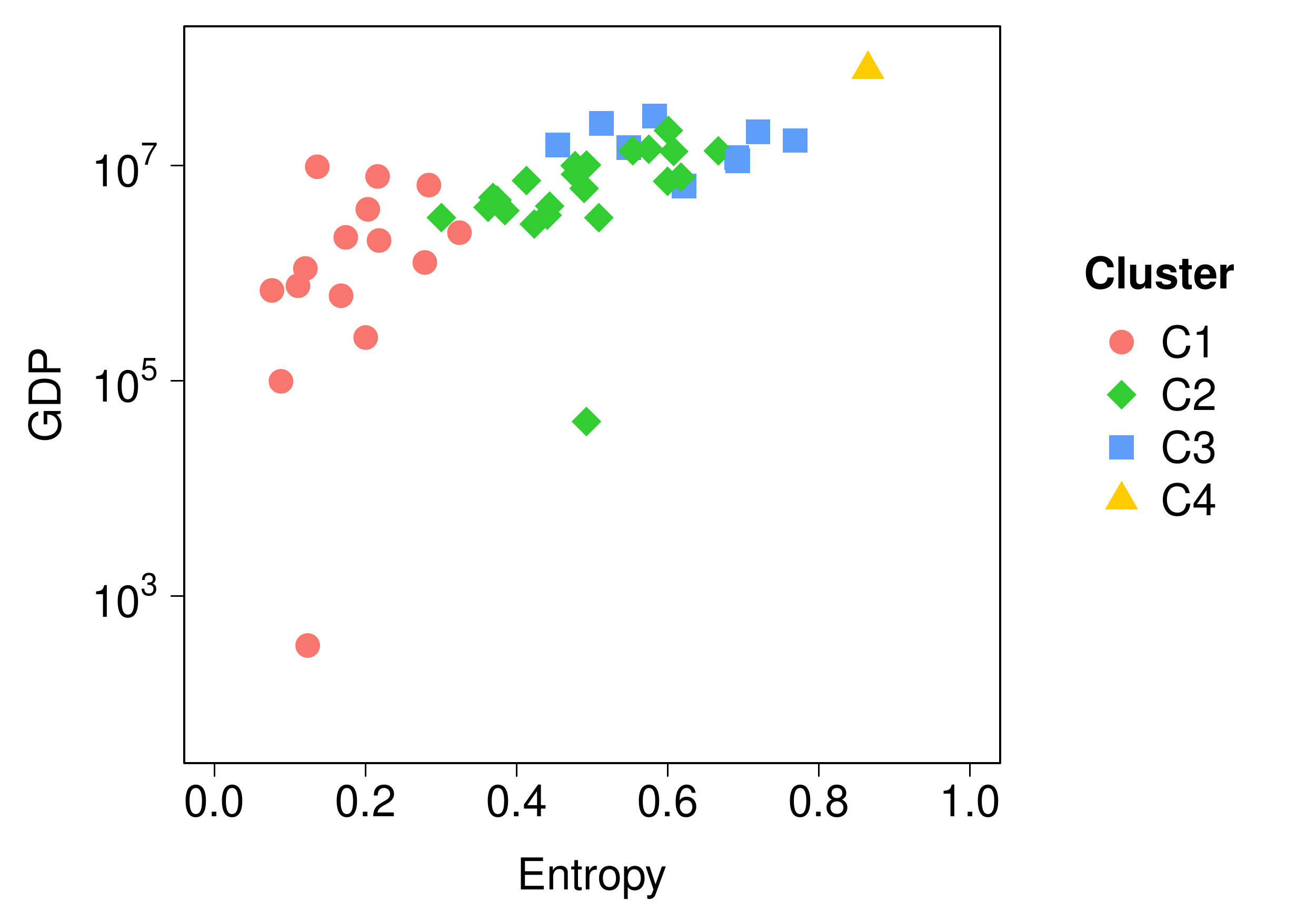}
	\caption{\textbf{Global Domestic Product (GDP) as a function of the entropy index.} The entropy have been averaged over the work shift time periods on weekdays.}
	\label{Socioeco1_SI}
\end{figure} 

\subsection*{Temporal evolution}

\begin{figure}[H]
	\centering
	\includegraphics[width=13cm]{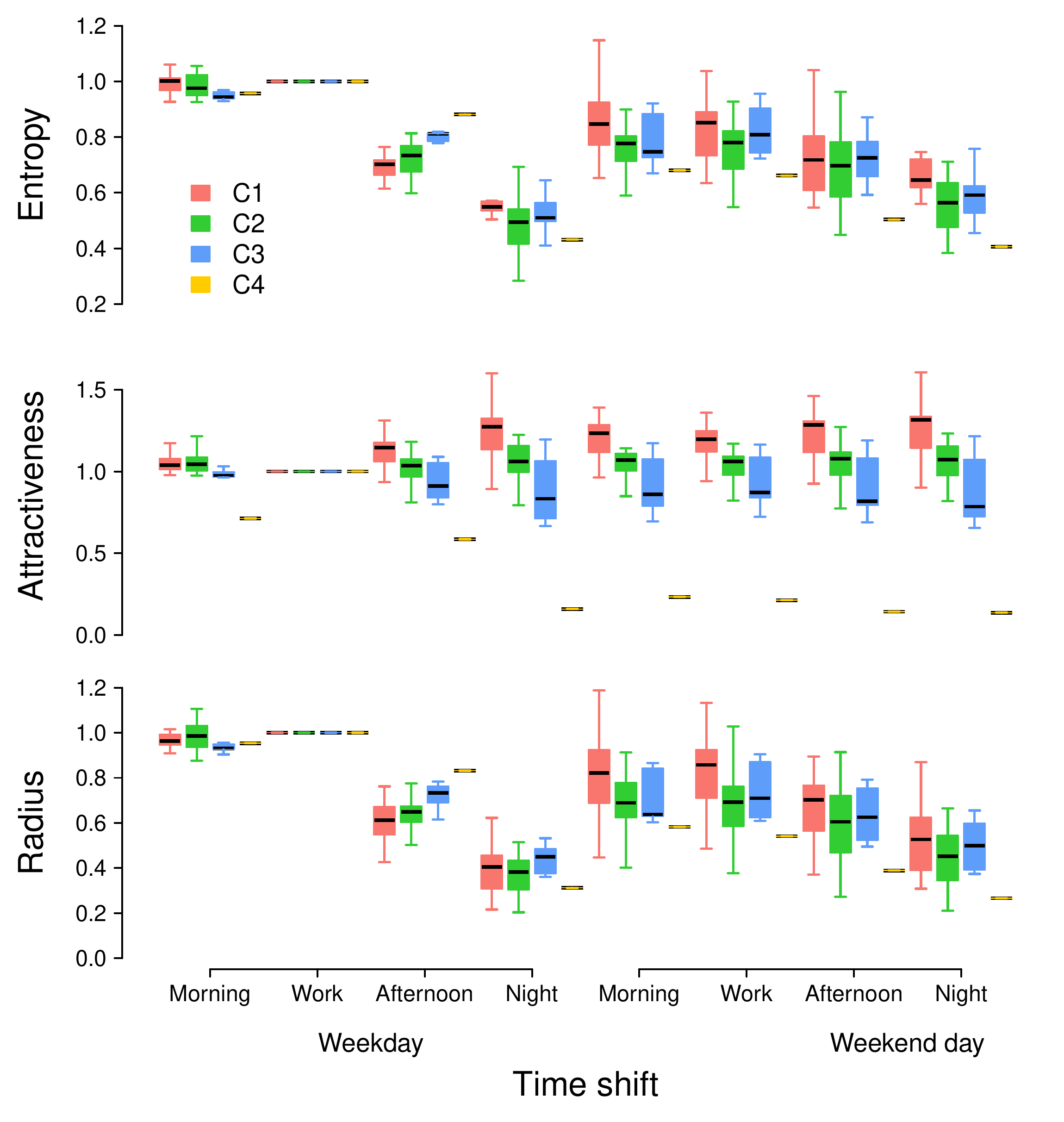}
	\caption{\textbf{Temporal evolution of the three metrics.} From the top to the bottom, Tukey boxplots of the entropy, attractiveness and radius of attraction as a function of time by cluster. The values are normalized by the value obtained for the work shift during weekdays.}
	\label{Time_SI}
\end{figure} 

\end{document}